\documentclass{PoS}
\let\ifpdf\relax
\usepackage{graphicx}
\usepackage{epsfig}
\usepackage{wrapfig}
\usepackage[small,bf]{caption}
\usepackage{amsmath}
\usepackage{amssymb}
\usepackage{dsfont}
\usepackage{slashed}
\usepackage{multirow}
\usepackage{lscape}
\usepackage{wrapfig}
\usepackage{MnSymbol}
\usepackage{adjustbox}
\usepackage{mciteplus}

\renewcommand{\d}{\textmd{d}}
\newcommand{\be}{\begin{equation}}
\newcommand{\ee}{\end{equation}}
\newcommand{\Tr}{\textmd{Tr}}
\newcommand{\tr}{\textmd{tr}}

\newcommand{\Z}{\mathcal{Z}}

\newcommand{\Spec}{\textmd{Spec}}

\newenvironment{myquote}{\list{}{\leftmargin=0.3in\rightmargin=0.3in}\item[]}{\endlist}

\title{QCD in magnetic fields:\\ from Hofstadter's butterfly to the phase diagram}
\ShortTitle{QCD in magnetic fields: from Hofstadter's butterfly to the phase diagram}
\date{\today}
\author{
Gergely Endr\H{o}di$^1$
}

\footnotetext[1]{Institute for Theoretical Physics, University of Regensburg, D-93040 Regensburg, Germany.}

\abstract {
I revisit the problem of a charged particle on a two-dimensional lattice immersed in 
a constant (electro)magnetic field, and discuss the energy spectrum -- Hofstadter's butterfly -- 
from a new, quantum field theoretical perspective. 
In particular, I point out that there is an intricate interplay between 
a) the structure of the butterfly at low magnetic flux, 
b) the absence of asymptotic freedom in QED and 
c) the enhancement of the quark condensate by a magnetic field at zero temperature.
I proceed to discuss the response of the QCD condensate to the magnetic field at nonzero 
temperatures in four space-time dimensions, present the resulting phase diagram and compare it to low-energy model predictions. }

\FullConference{The XXXII International Symposium on Lattice Field Theory \\
                June 23 - 28, 2014\\
                New York, NY, USA.}

\begin{document}

\section{Acknowledgments}

It is a great honor to be selected to receive this year's Ken Wilson prize for 
{\it significant contributions to our understanding of
QCD matter in strong magnetic fields and to QCD
thermodynamics}. 
This recognition urges me to keep up doing good research and to contribute 
to our field as significantly as I possibly can. 
I have not been so lucky to meet Kenneth G.\ Wilson in person, but reading through 
a collection of anecdotes~\cite{Kronfeld:2013lda,*2014arXiv1407.1855G}, I feel that I would 
have greatly enjoyed interacting with him.
One of his quotes I find particularly appropriate 
and useful as a guideline:

\begin{myquote}
\it
You shouldn't choose a problem on the basis of the tool. You start by thinking
about the physics problem \ldots
maybe you'll solve it using computer techniques, maybe using a contour
integral; but it's very important to approach it starting from the physics because
otherwise you get lost in the use of the tool, and lose track of where you're trying
to go.
\end{myquote}

\noindent
Although most problems in lattice gauge theory are more likely to 
be solved by a supercomputer than by contour integration, I still believe 
it is important to bear this principle in mind. 
This is one of the very reasons I find QCD in magnetic fields so fascinating: 
besides the problems requiring large-scale numerical simulations, 
various aspects of the topic also allow for an analytical treatment.
We will see examples for both during the talk.

At this point I would like to express my gratitude to my collaborators; 
this research would not have been possible without their help. 
I would especially like to thank my colleagues and friends with whom I collaborated on 
magnetic field-related topics: Gunnar Bali, Falk Bruckmann, Martha Constantinou, Marios Costa, 
Zolt\'an Fodor, S\'andor Katz, Tam\'as Kov\'acs, Stefan Krieg, Haris Panagopoulos, 
Andreas Sch\"afer and K\'alm\'an Szab\'o. 
I am also grateful for many enlightening discussions with Jens Oluf Andersen,
Szabolcs Bors\'anyi, Pavel Buividovich, Massimo D'Elia, Gerald Dunne, 
Eduardo Fraga, Christof Gattringer, Antal Jakov\'ac, 
Claudia Ratti, Marco Ruggieri, Hans-Peter Schadler, Andreas Schmitt, Igor Shovkovy 
and B\'alint T\'oth.

\section{Introduction: butterflies and lattices}

The physics of Quantum Chromodynamics (QCD) in the presence of background magnetic fields 
is remarkably rich. 
The most interesting aspects are how 
the magnetic field breaks rotational symmetry inducing 
anisotropic pressures; how it affects chiral symmetry breaking and deconfinement; 
and how it modifies various hadronic properties. 
Besides these challenging theoretical
concepts, background magnetic fields in QCD 
have various applications including the cosmology
of the early universe, 
non-central heavy-ion collisions and magnetized neutron stars~\cite{Kharzeev:2013jha}.

\begin{wrapfigure}{r}{7.cm}
 \vspace*{-.1cm}
 \centering
\includegraphics[width=6.8cm]{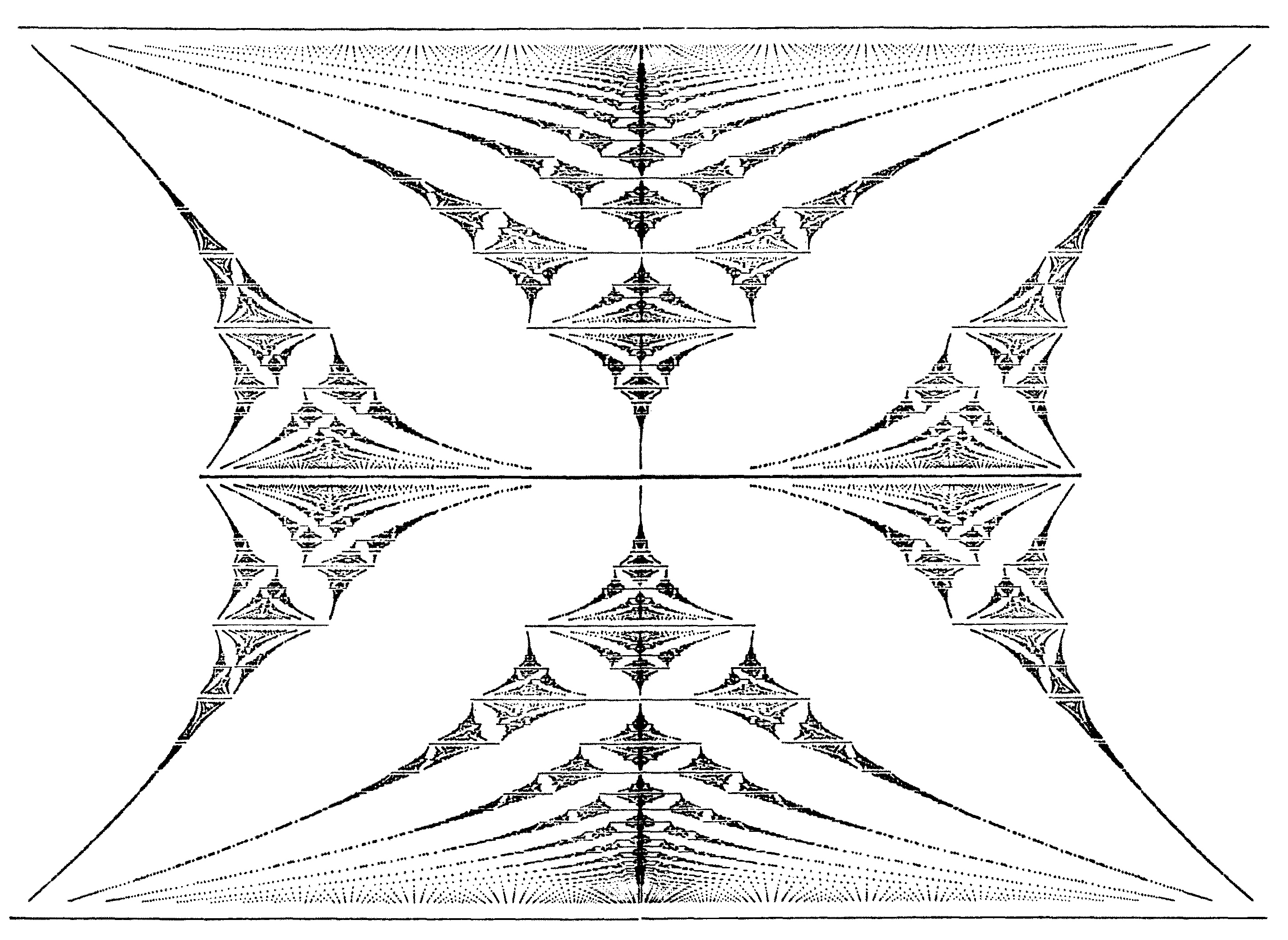}
\vspace*{-.1cm}
 \caption{\label{fig:origbutt}
 Hofstadter's butterfly~\protect\cite{Hofstadter:1976zz}.
}
 \vspace*{-.1cm}
\end{wrapfigure}

A particularly beautiful example for the complexity that magnetic fields can induce is 
provided by Hofstadter's butterfly~\cite{Hofstadter:1976zz}: the energy spectrum of a 
Bloch electron immersed in a background magnetic field, see Fig.~\ref{fig:origbutt}. 
The plot summarizes the quantum mechanically allowed energy levels (on the horizontal 
axis) for various values of the magnetic flux (on the vertical axis).
The butterfly exhibits an apparently recursive hierarchy, with the coarse structure 
repeated on finer levels in ever smaller copies.
A hint towards understanding this recursive pattern is provided by considering 
two aspects of the problem separately: the interaction of the electron with the 
infinite periodic lattice potential
(with lattice spacing $a$), 
and its coupling to the background magnetic field $B$. 
The former 
problem gives rise to periodic Bloch waves, 
whereas the latter is described in terms of Landau levels. 
Both
aspects involve a typical frequency~\cite{Hofstadter:1976zz}: In the non-relativistic treatment 
of a particle with mass $m$ and charge $q$, 
Landau levels are characterized by the cyclotron frequency $qB/m$, while the 
frequency of the Bloch wave with maximal momentum is $(2\pi/a) / (ma)$. The 
ratio of these two characteristic frequencies equals
\be
\alpha\equiv a^2qB/(2\pi),
\label{eq:Halpha}
\ee
which is 
proportional to the magnetic flux through 
an elementary plaquette of the lattice.
It turns out that if the characteristic frequencies are commensurable (i.e., if 
$\alpha\in\mathds{Q}$), the energy eigenvalues accumulate into finite bands, 
separated by finite gaps.
In contrast, if the involved frequencies are incommensurable ($\alpha\not\in\mathds{Q}$) 
the energy spectrum dissolves into a zero-measure nowhere dense set that is isomorphic 
to the Cantor set~\cite{Hofstadter:1976zz,last}. In some sense the two problems are in 
this case incompatible with each other, and the particle becomes unable to obey both the 
periodic structure of the Bloch wave {\it and} the circular structure of the Landau levels.  
This frustration manifests itself in the fractal structure of the butterfly.

In the present talk, this solid state physics problem will be interpreted from a new, 
quantum field theoretical (QFT) point of view. Specifically, I will consider the QCD vacuum 
exposed to background magnetic fields 
in the lattice regularization. As we will see, the eigenvalue spectrum of this 
problem, for vanishing gauge coupling in two dimensions, coincides with the energy levels 
of the solid state physics example. 
On the one hand, the original `solid state physics butterfly' 
is obtained in terms of a physical lattice spacing, characteristic to the crystal on which 
the electron lives. On the other hand, for the `QFT butterfly' the lattice spacing 
plays the role of a regulator. This regulator (after renormalization is performed) 
is removed from the theory via the continuum limit $a\to0$. In this limit 
the lattice structure disappears, which, in turn, implies that the `QFT butterfly' 
is merely a lattice artefact. Still, as will be argued below, certain aspects of the 
butterfly do survive the $a\to0$ limit and correspond to concepts in continuum physics. 
After pointing out these aspects, I will generalize the discussion 
to full dynamical QCD in four space-time dimensions, and discuss the phase diagram 
in the magnetic field - temperature plane using non-perturbative lattice simulations.

\section{Magnetic field, spectrum and symmetries}

\subsection{In the continuum}

The interaction with the background magnetic field proceeds via minimal coupling, 
with a Landau-gauge electromagnetic potential $A_\mu$,
\be
D_\mu = \partial_\mu+iq A_\mu, \quad\quad\quad A_y=Bx, \quad\quad A_\nu=0, \quad \nu\neq y,
\label{eq:Acont}
\ee
where $q$ denotes the charge of the particle and the coordinate system was oriented such that the magnetic field points in the 
positive $z$ direction. 
For charged bosons, the equation of motion involves the Klein-Gordon operator $D^2$, 
whereas for charged fermions, the Dirac operator $\slashed{D}=\gamma_\mu D_\mu$, 
where $\gamma_\mu$ denote the Euclidean $\gamma$-matrices. 
Besides the coupling to the magnetic field, there are no other interactions: this 
is what I will refer to below as the `free' case.

The eigenvalues of $\sqrt{-D^2}$ and $i\slashed{D}$, respectively, give the allowed energy 
levels in the bosonic and fermionic problems. Simple $\gamma$-matrix identities 
show that the two operators are related to each other via
\be
\slashed{D}\,{}^2 = D^2 + \sigma \cdot qB,\quad\quad\quad
\sigma \equiv \frac{1}{2i}\,[\gamma_x,\gamma_y].
\label{eq:DD}
\ee
The three operators $\slashed{D}{}^2$, $D^2$ and $\sigma$ pairwise commute with each other, 
and thus have common eigenvectors.
Eq.~(\ref{eq:DD}) acting on each joint eigenvector produces a relation 
between the eigenvalues, written compactly as
\be
\Spec\{\slashed{D}{}^2\} = \Spec\{D^2\} + \Spec\{\sigma\} \cdot qB.
\label{eq:DDD}
\ee

The eigenvalues of all three operators can be found in a straightforward way. 
The solutions are written in terms of Landau levels indexed by a non-negative integer $n$ and 
the component $s=\pm1/2$ of the spin in the direction of the magnetic field. 
They indeed fulfill Eq.~(\ref{eq:DDD}):
\be
\Spec\{\slashed{D}^2\} = -2qB(n+1/2+s), \quad\quad \Spec\{D^2\} = -2qB(n+1/2), \quad\quad 
\Spec\{\sigma\} = 2s.
\label{eq:Landau}
\ee

\subsection{On the lattice}

On the lattice, the gauge potential~(\ref{eq:Acont}) is implemented through $\mathrm{U}(1)$ phases 
$u_\mu(n)$ that live on the links between lattice sites,
\be
u_y(n)=\exp(ia^2q B \,n_x), \quad\quad u_x(n) = \exp(-ia^2qB\,N_x n_y \cdot \delta_{n_x,N_x-1}),
\quad\quad u_z(n)=u_t(n)=1,
\label{eq:links}
\ee
where the sites are labeled by integers $n=(n_x , n_y , n_z , n_t )$, with 
$n_\mu=0\ldots N_\mu-1$ and $a$ is the lattice spacing. For convenience, the lattice extents 
$N_\mu$ are taken to be even integers. 
The `twist' of the $x$-links at $n_x=N_x-1$ is necessary to satisfy periodic boundary 
conditions for the gauge potential and to ensure the constancy of the magnetic field throughout the 
$x-y$ plane~\cite{Martinelli:1982cb}. 
In this setup, the flux of the magnetic field is quantized~\cite{'tHooft:1979uj},
\be
qB\cdot a^2N_xN_y = 2\pi N_b, \quad\quad N_b\in \mathds{Z}, \quad\quad 0\le N_b <N_xN_y.
\ee
Thus, there is a minimal magnetic field due to the finiteness of the volume. 
In addition, due to the periodicity of the links~(\ref{eq:links}) in $N_b$, 
there is also a maximal possible magnetic field, which is set by the square of the 
inverse lattice spacing.
In terms of $N_b$, Hofstadter's parameter~(\ref{eq:Halpha}) reads
\be
\alpha = N_b/ ( N_xN_y), \quad\quad\quad 0\le\alpha\le 1.
\ee
This implies that on a finite lattice $\alpha$ is always rational, and the true fractal 
butterfly only emerges in the infinite volume limit $N_xN_y\to\infty$. In this limit, 
an irrational value of $\alpha$ is obtained as the ratio of two integers, both 
of which approach infinity. Similarly, the continuous bands at 
$\alpha\in\mathds{Q}$ are only present in the infinite volume limit and are
approached by the discrete spectra on finite lattices.

Next, we have to specify the lattice discretization of the operator appearing in 
the equation of motion. For the bosonic case, the simplest discretization is
\be
D^2_{nm} = \frac{1}{a^2}\sum_{\mu} \left[ u_\mu(n)\,\delta_{m,n+a\hat\mu} 
+ u_\mu^\dagger(n-a\hat\mu)\, \delta_{m,n-a\hat\mu}  - 2 \delta_{m,n}\right].
\label{eq:Db}
\ee
In the fermionic case, the most convenient discretization is the staggered 
formulation. Here, the $\gamma$-matrices are replaced by space-dependent phases $\eta_\mu$, which 
result from a local transformation of the fermion field that diagonalizes the action 
in spinor space. 
The resulting lattice Dirac operator, again in the presence of the $\mathrm{U}(1)$ phases 
Eq.~(\ref{eq:links}) is written as
\be
\slashed{D}_{nm} = \frac{1}{2a} \sum_{\mu} \left[ u_\mu(n) \,\eta_\mu(n) \,\delta_{m,n+a\hat\mu} 
- u_\mu^\dagger(n-a\hat\mu)\,\eta_\mu(n-a\hat\mu)\, \delta_{m,n-a\hat\mu} \right], 
\quad\quad\quad
\eta_\mu(n) = (-1)^{\sum_{\nu=x}^{\mu-1} n_\nu}.
\label{eq:Df}
\ee
In the following, the discussion will be restricted to two spatial dimensions, such 
that the sums in Eqs.~(\ref{eq:Db}) and~(\ref{eq:Df}) include $\mu=x$ and $y$. 

\begin{figure}[ht!]
 \centering
 \vspace*{-.2cm}
 \mbox{
\includegraphics[width=7.4cm]{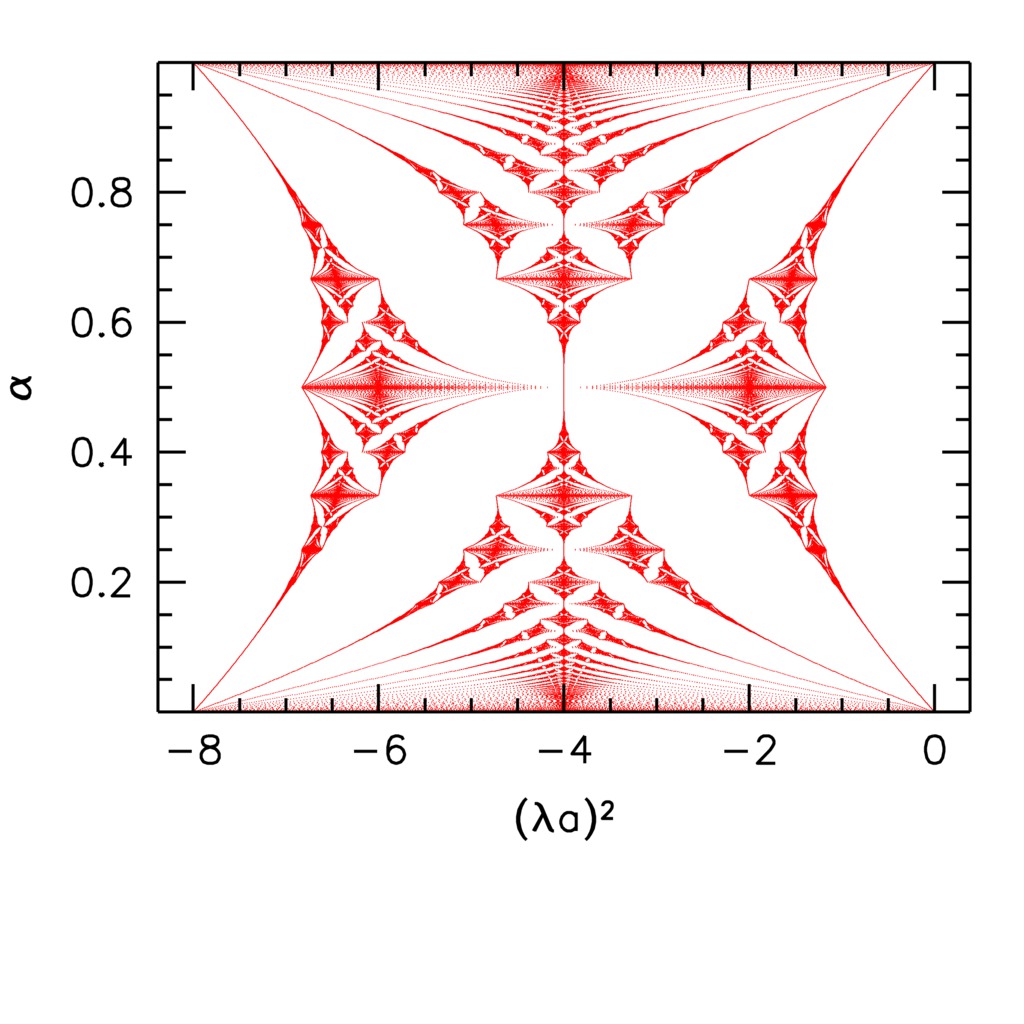} \quad
 \includegraphics[width=7.4cm]{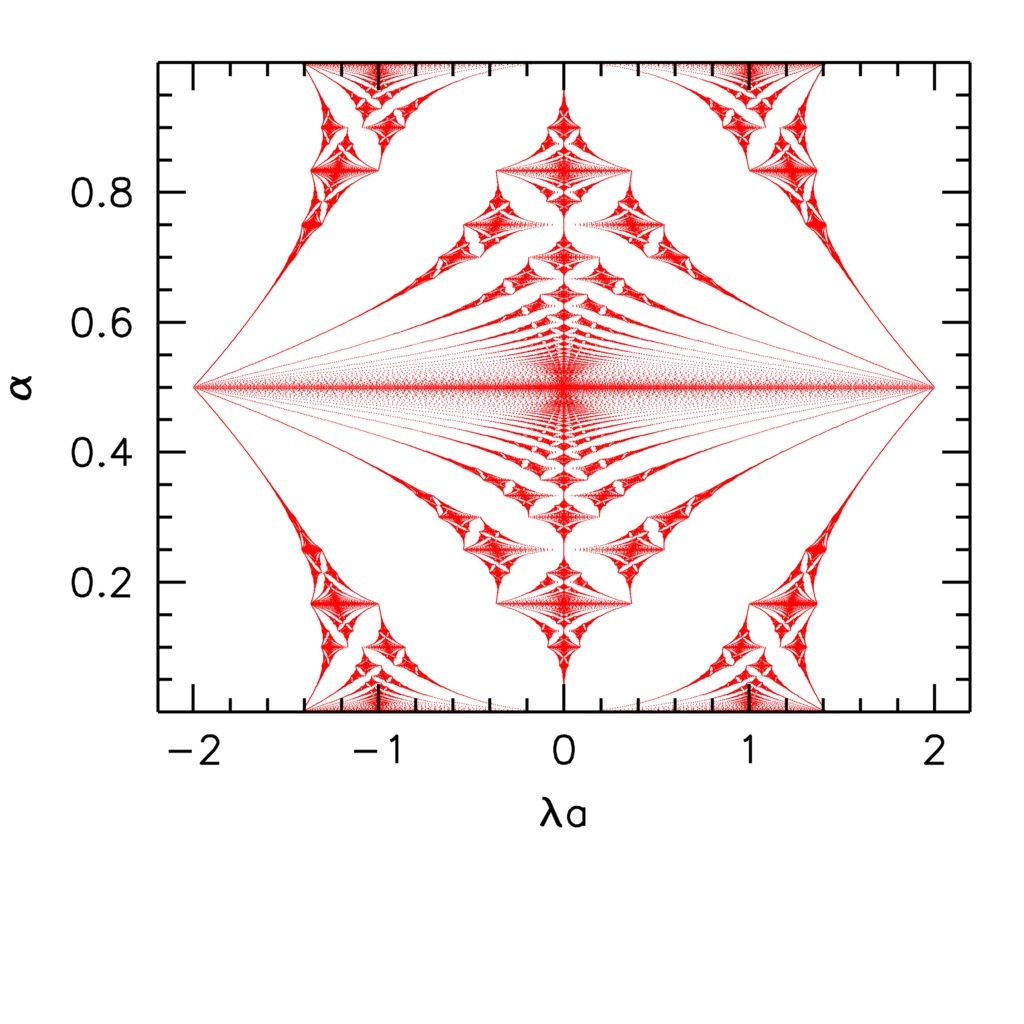}}
\vspace*{-2cm}
 \caption{\label{fig:butterflies}
 Butterfly in the bosonic (left panel) and fermionic (right panel) cases. 
 The eigenvalues in lattice units are plotted against the strength of the 
 magnetic field.}
\end{figure}

The lattice eigenvalues of $i\slashed{D}\,$ and those of $D^2$ can be found numerically 
by diagonalizing 
the corresponding matrices (note that both of these operators are Hermitean 
and thus have a real spectrum).
A lattice of size $N_x=N_y=40$ is considered 
and the Lapack library is used to determine all eigenvalues. The results 
are shown in Fig.~\ref{fig:butterflies} for the bosonic and fermionic cases. 
Apparently, the two spectra are related to each other by a simple transformation (a 
translation by four and a rescaling by two) as well as by a shift in the magnetic 
field parameter $\alpha$,
\be
\Spec\{D^2(\alpha)\} = 2\cdot\Spec\{i\slashed{D}((\alpha+1/2) \textmd{ mod } 1)\} - 4.
\label{eq:symmfb}
\ee
This relation can be proven analytically by exploiting the exact form of the 
operators~(\ref{eq:Db}) and~(\ref{eq:Df}) 
after inserting the links~(\ref{eq:links}) 
and using the fact that $N_x$ is even.
Specifically, the staggered phases $\eta_x=1$, $\eta_y=\exp(i\pi n_x)$ 
induce the shift in the magnetic field by one half of the period, cf.\ Eq.~(\ref{eq:links}). 
The sign in front of the adjoint links 
(positive in the bosonic and negative in the fermionic case) can be gauged 
away and does not modify the spectrum\footnote{
The corresponding local $\mathrm{U}(1)$ gauge transformation reads
\be
\psi (n) \to i^{\,n_x+n_y} \psi(n),
\ee
which effectively multiplies all links in the $x-y$ plane by $i$ and, thus, 
flips the sign of the adjoint links and produces an overall factor $i$. 
Note that this is only true if $N_x$ and $N_y$ are both multiples of four. 
This constraint, however, is a boundary effect, which disappears in the 
infinite volume limit. A similar argument was discussed in Ref.~\cite{2012arXiv1210.6355K}. 
}. 
Further symmetries of the spectra are
\be
\Spec\{i\slashed{D}\,(\alpha)\} = \Spec\{i\slashed{D}(1-\alpha)\}, \quad\quad\quad 
\Spec\{i\slashed{D}(\alpha)\} = \Spec\{-i\slashed{D}(\alpha)\},
\label{eq:symmr}
\ee
which follow from the periodicity of the $\mathrm{U}(1)$ links in $\alpha$, 
parity symmetry and chiral symmetry.
The operator discussed by Hofstadter~\cite{Hofstadter:1976zz} 
was $D^2$, i.e.\ the original butterfly\footnote{
Even though the feedback of several people in the audience showed that the 
fermionic spectrum does not resemble an actual butterfly, 
in the absence of a spot-on alternative I will stick to this nomenclature. 
} 
actually describes scalar particles and 
corresponds to the spectrum in the left panel of Fig.~\ref{fig:butterflies} (cf.\ Fig.~\ref{fig:origbutt}). Nevertheless, through the symmetry~(\ref{eq:symmfb}), 
the fermionic and bosonic spectra are in a one-to-one correspondence to each other. 
It is important to mention that the butterfly is (to a certain extent) also accessible 
in experiments: 
the electrical conductivity of graphene samples was 
found to exhibit a fractal pattern in magnetic fields~\cite{2013Natur.497..598D,*2013Natur.497..594P}. 

\section{Correspondence between continuum and lattice}

Let me proceed by pointing out the differences and similarities between the 
continuum and lattice spectra. As was mentioned in the introduction, in the 
quantum field theoretical setting the lattice spacing is a mere regulator 
that is eliminated in the continuum limit. Correspondingly, the butterfly 
must disappear as $a\to0$. Indeed, at fixed magnetic field $B$ the parameter 
$\alpha$ approaches zero in this limit and, thus, only the low-$\alpha$ end of the spectrum 
may play a physical role. Therefore, it is instructive to compare the lattice and 
continuum settings in this region.

\begin{wrapfigure}{r}{7.6cm}
 \centering
\vspace*{-.5cm}
 \includegraphics[width=7.3cm]{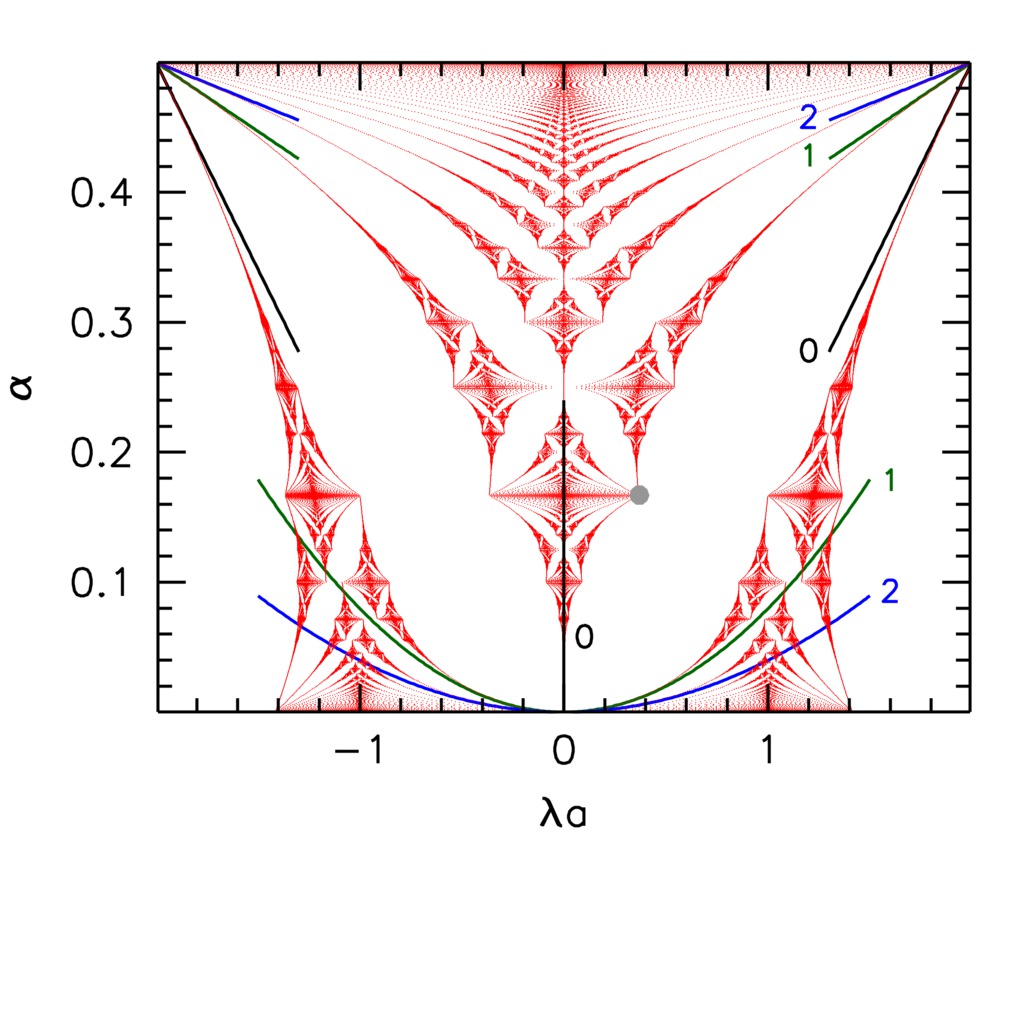}
\vspace*{-1.4cm}
 \caption{\label{fig:laby}The butterfly as a labyrinth of fermionic (solid lines 
starting at the origin) and bosonic (solid lines starting at the upper corners) continuum 
Landau levels. 
 The finer structure is generated by recursively appearing 
 smaller copies of this skeleton -- see a zeroth fermionic Landau level starting at the 
gray dot.
}
\vspace*{-.5cm}
\end{wrapfigure}

In the continuum, 
the energy eigenvalues obey the Landau-level structure Eq.~(\ref{eq:Landau}). In particular, 
the lowest levels are given by
\be
\textmd{fermions:}\quad
(\lambda a)^2 = qB \cdot (0,2,\ldots), \quad\quad\textmd{bosons:}\quad
(\lambda a)^2 = qB\cdot(1,3,\ldots).
\label{eq:contland}
\ee
In the low-$\alpha$ region, the lattice eigenvalues indeed 
follow these continuum energies, see 
Fig.~\ref{fig:laby}, where the lower half of the fermionic butterfly is shown. 
The fermionic Landau levels are the curves starting from the origin in the plot.
In addition, combining the symmetries~(\ref{eq:symmfb}) and~(\ref{eq:symmr}) also 
reveals 
that the vertical reflection of the fermionic butterfly coincides (up to a 
simple 
linear rescaling) with the bosonic butterfly. Therefore, the bosonic Landau levels 
also show up in the same diagram, starting at the upper corners of Fig.~\ref{fig:laby}. 
Notice that while the fermionic levels 
(the eigenvalues of $i\slashed{D}$) are proportional to 
$\sqrt{qB}$, the bosonic levels (the eigenvalues of $D^2$) are proportional to $qB$. As a result, 
the fermionic curves are quadratic, while the bosonic ones linear in the figure. Note moreover 
that the lowest fermionic Landau level is independent of $B$. 

As the magnetic flux increases, the lattice eigenvalues tend to deviate from the continuum curves, 
dissolve into bands and mix with each other, forming the recursive pattern. 
Notice that this breaking up of the continuum curves proceeds in an 
apparently similar fashion for all Landau levels. 
Therefore, the `coarsest' structure of the butterfly is provided by continuum Landau levels, 
which are all similar 
to each other and also get repeated on the finer levels. Notice, for example, a sub-structure zeroth Landau level starting 
in the middle of the figure, indicated by the gray dot.
This is a new (qualitative) representation of Hofstadter's butterfly as a labyrinth of 
hierarchically embedded fermionic and bosonic Landau levels.

\subsection{The quark condensate}
\label{sec:qc}

\begin{figure}[b]
 \centering
\vspace*{-.7cm}
\mbox{
 \includegraphics[width=7.4cm]{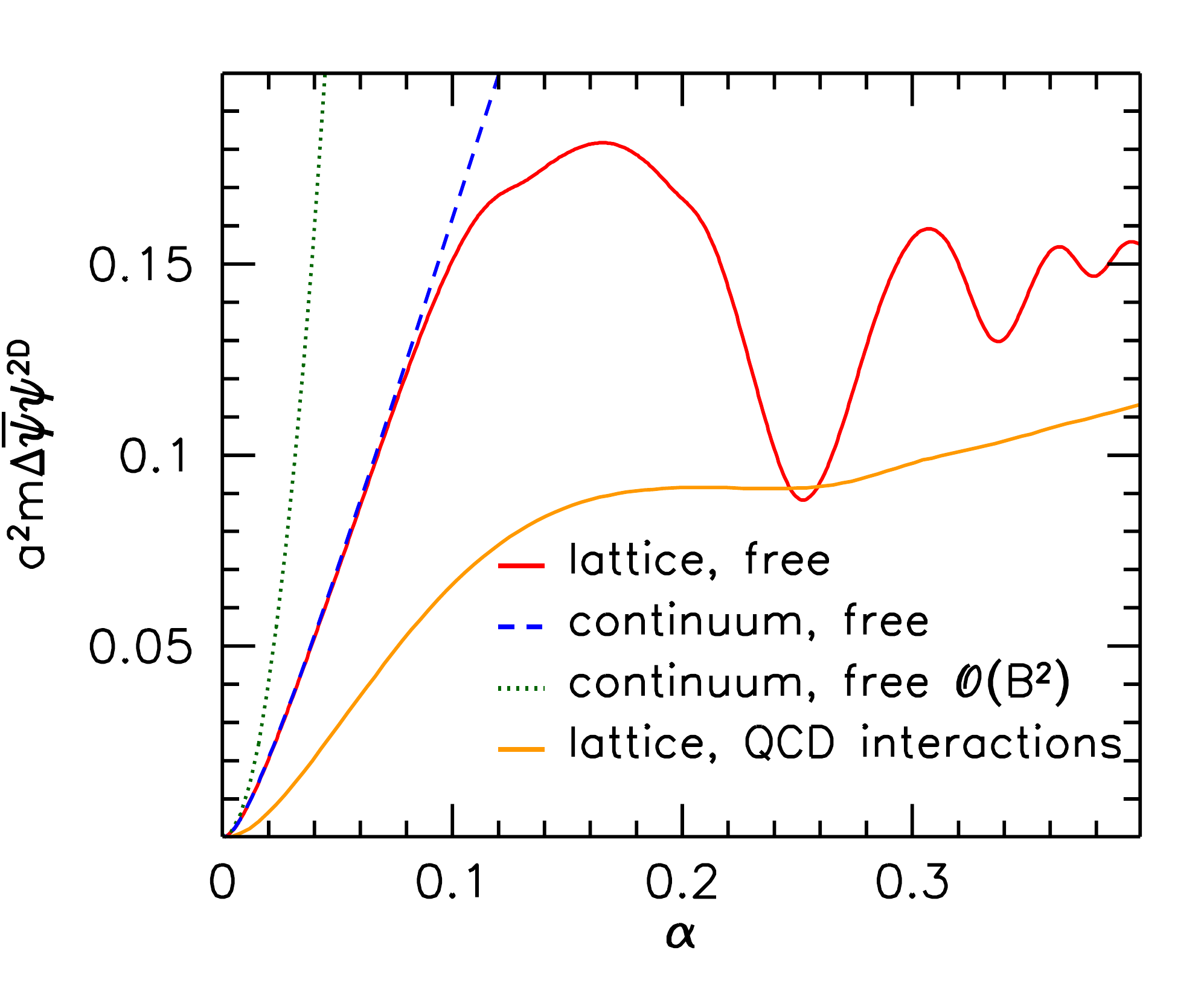} \quad
 \includegraphics[width=7.4cm]{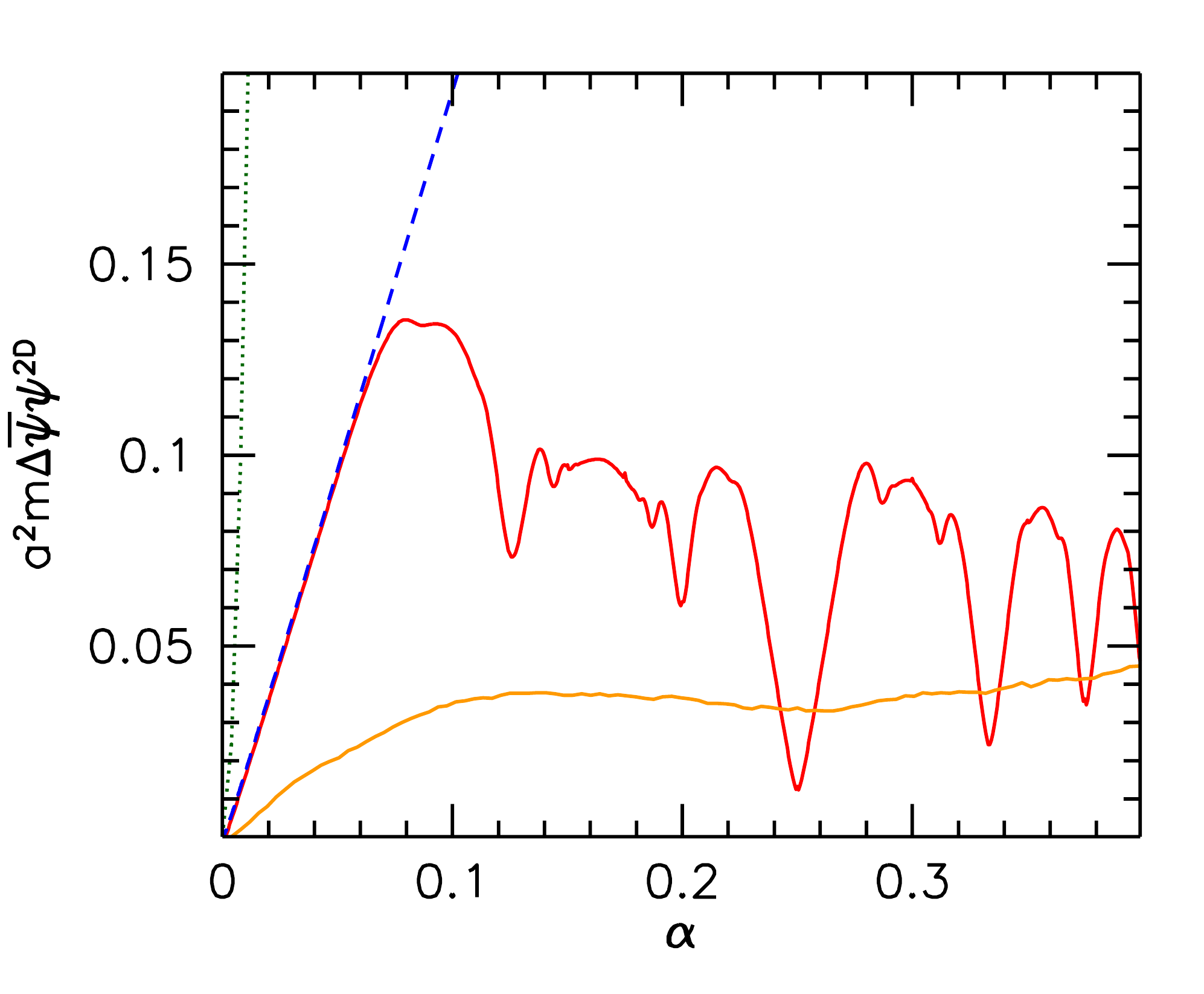} }
\vspace*{-.8cm}
 \caption{\label{fig:pbp}The quark condensate on the lattice (solid red line) at two 
intermediate values of the mass ($ma=0.2048$ in the left panel and $ma=0.0512$ in the right panel). Also indicated are the continuum condensate (blue dashed line), 
the quadratic contribution to $\Delta\bar\psi\psi^{\rm 2D}$ (green dotted line) 
and the condensate calculated on the lattice in the presence of QCD interactions (solid orange line).} 
\vspace*{-.5cm}
\end{figure}

Besides the difference of the role played by the lattice spacing,
there is another important difference between the `solid state butterfly' and the `QFT butterfly'.
Namely, the energy of a single electron is in principle measureable, while
one eigenvalue of $i\slashed{D}$ has no physical meaning. Instead, physical observables are typically obtained by 
combining all eigenvalues of the Dirac operator into a spectral sum. One of the most important observables in QCD 
is the quark condensate, which can be written in such a spectral representation as
\be
\bar\psi\psi^{\rm 2D} \equiv 
\frac{1}{V_2} \tr (\slashed{D}\,+m)^{-1} = 
\frac{1}{V_2} \sum_j \frac{m}{\lambda_j^2 + m^2},
\label{eq:pbpdef}
\ee
where $m$ denotes the mass of the quark and $V_2=a^2N_xN_y$ the two-dimensional 
volume. In addition, the change of the condensate induced by the magnetic 
field is defined as
\be
\Delta \bar\psi\psi^{\rm 2D} \equiv \left.\bar\psi\psi^{\rm 2D}\right|_{B} - 
\left.\bar\psi\psi^{\rm 2D}\right|_{B=0}.
\ee
For each value of the magnetic field (i.e.\ for each $\alpha$) the condensate 
is calculated from the corresponding lattice eigenvalues according to Eq.~(\ref{eq:pbpdef}). 
The so obtained condensate difference is plotted in Fig.~\ref{fig:pbp} for two intermediate values of the mass in the low-$\alpha$ region.
At masses much larger than the typical differences between the eigenvalues, 
the dependence of $\Delta\bar\psi\psi^{\rm 2D}$ on $\alpha$ is completely smooth, since 
the mass washes out the irregular changes in the eigenvalues as $\alpha$ is tuned. 
As $m$ is reduced, more and more of the fractal pattern of the butterfly 
becomes visible, revealed by oscillatory segments that appear in the condensate. 
In the chiral limit the oscillations approach a fractal curve that fully explores 
the recursive structure of the butterfly. 
A simple demonstration that was presented during the talk to illustrate the 
dependence of 
$\Delta\bar\psi\psi^{\rm 2D}$ on the mass and on the magnetic field can be found at 
the url~\cite{demo}.

The condensate can also be calculated using the continuum 
energies Eq.~(\ref{eq:Landau}). The result -- see Eq.~(\ref{eq:pbpproper}) in 
App.~\ref{sec:app1} -- is indicated in Fig.~\ref{fig:pbp} by the blue dashed lines. 
The agreement 
between the lattice and the continuum condensates at low magnetic fields again 
shows that the low-$\alpha$ part of the Hofstadter spectrum represents continuum physics. 
In fact, irrespective of how low the mass is, the initial segment of the 
lattice condensate is always smooth and follows the dependence dictated by 
the continuum Landau levels.
The leading-order dependence of the condensate on the magnetic field (for 
non-vanishing masses) is quadratic in $B$. The corresponding coefficient 
-- as demonstrated by explicit calculation in App.~\ref{sec:app1} -- 
is proportional to the lowest-order coefficient $\beta_1=1/(12\pi^2)$ 
of the $\beta$-function of four-dimensional quantum electrodynamics (QED),
\be
\Delta \bar\psi\psi^{\rm 2D} = (qB)^2 \cdot \beta_1 \cdot \frac{4\pi}{m^3} + \mathcal{O}(B^4).
\label{eq:pbp1cloned}
\ee
The sign of $\beta_1$ is fixed by the leading renormalization group behavior 
of the theory: the absence of asymptotic freedom in QED ensures that $\beta_1>0$. 
In turn, through Eq.~(\ref{eq:pbp1cloned}) this results in a quadratic increase 
of the condensate with growing magnetic field. 
This observation leads to the following, three-fold correspondence:\\[.1cm]
\be
\raisebox{-25pt}{\hspace*{.5cm}\includegraphics[width=13.5cm]{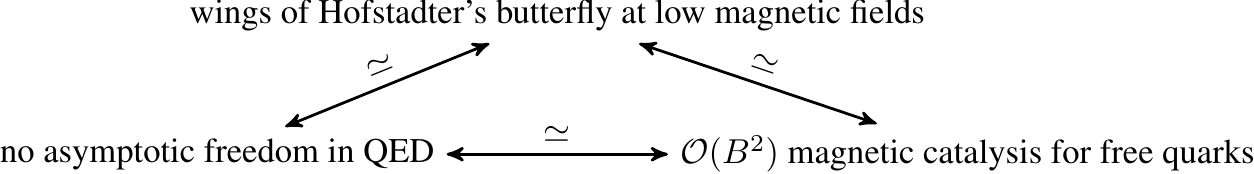}}
\vspace*{.4cm}
\label{eq:correspondence}
\ee
where `magnetic catalysis'~\cite{Gusynin:1995nb} refers to the enhancement of the 
condensate by $B$.
I find this correspondence quite remarkable, as it connects three, seemingly
unrelated phenomena: the spectrum 
in a solid state physics problem, a notion in 
perturbative quantum field theory and a characteristic about the breaking 
of chiral symmetry by the magnetic field. 

\begin{wrapfigure}{r}{7.5cm}
 \centering
 \vspace*{-.5cm}
\includegraphics[width=7.8cm]{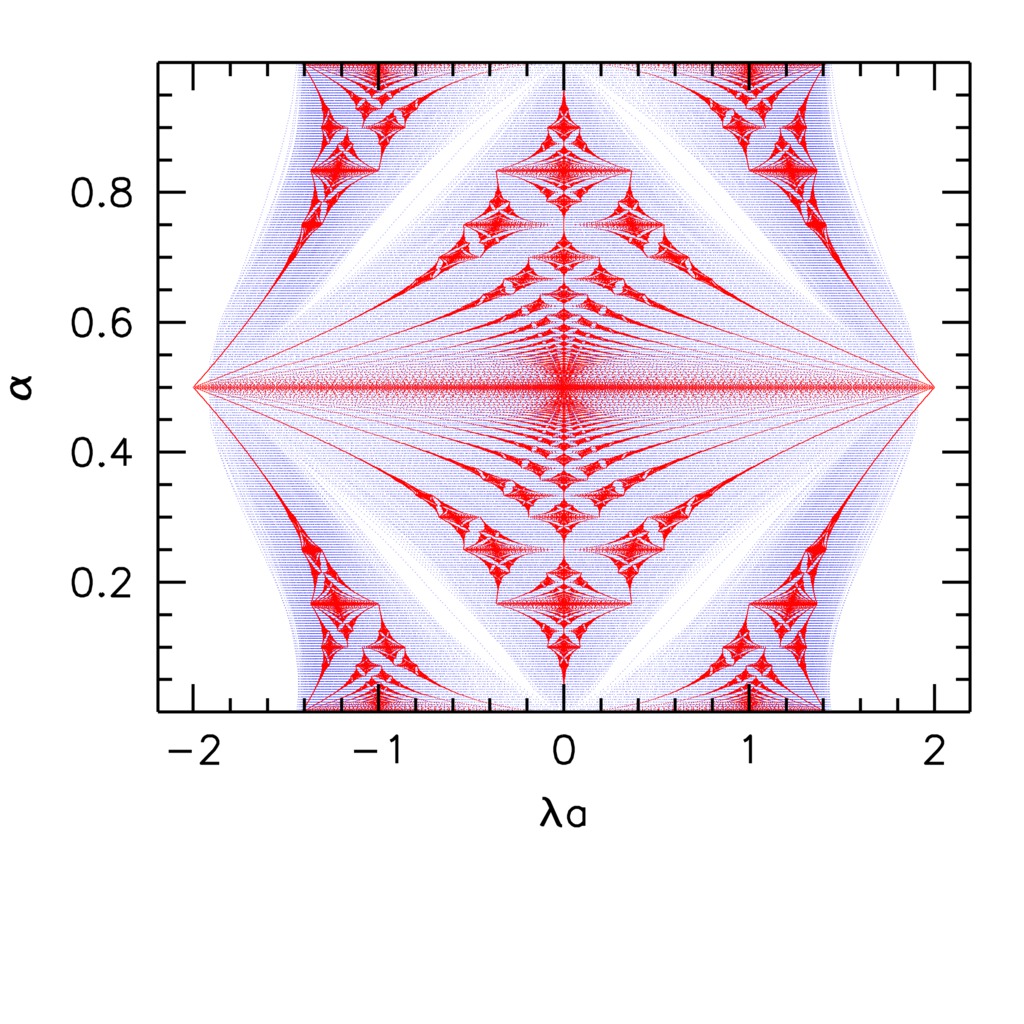}
\vspace*{-1.9cm}
 \caption{\label{fig:butti} Fermionic butterfly in the free case (red) and 
with QCD interactions switched on (blue). }
\vspace*{-.6cm}
\end{wrapfigure}

Up to this point, the properties of the non-interacting Dirac operator were discussed. 
Let me 
now continue by making contact to the case of full QCD, considering the effect of 
gluonic interactions on the two-dimensional spectrum. In the presence 
of gluons, the $\mathrm{U}(1)$ links $u_\mu$ in the Dirac operator (\ref{eq:Df}) are 
promoted to 
$\mathrm{SU}(3)\times \mathrm{U}(1)$ matrices $U_\mu\cdot u_\mu$. 
The interacting operator is diagonalized on an $x-y$ plane of a typical four-dimensional gluonic 
configuration. The resulting spectrum -- together with the fermionic butterfly of the free 
case -- is shown in Fig.~\ref{fig:butti}. Apparently, QCD interactions tend to wash out the 
fractal structure and the Lan-

\noindent
dau levels composing the wings of the butterfly get smeared out. 
Nevertheless, 
the coarse structure of the spectrum -- in particular the gap between the zeroth and first fermionic 
Landau levels -- remains present in the interacting case as well. 
This qualitative similarity shows up in the dependence of the condensate on the magnetic field as well. 
Fig.~\ref{fig:pbp} also includes $\Delta\bar\psi\psi^{\rm 2D}$ in the interacting case, featuring the quadratic increase of the condensate at low magnetic fields and the damped oscillations at larger 
values of $\alpha$.

As a side remark, I mention that the Fourier transform of the condensate 
with respect to $\alpha$ gives 
the so-called dual condensate, which can be used to 
calculate Wilson loops of fixed area~\cite{Bruckmann:2011zx}, and thus 
makes contact to confinement in the interacting QCD case.

\section{Magnetic catalysis, inverse catalysis and the phase diagram}

Above I argued that the quadratic increase of the condensate of free quarks 
at low magnetic fields is related to the positivity of the QED 
$\beta$-function. (As the calculation in App.~\ref{sec:app1} shows, this 
holds both in two and in four dimensions.)
The results for the condensate of interacting quarks (see Fig.~\ref{fig:pbp}) 
seem to indicate that 
the quadratic enhancement by $B$ is exhibited in the presence of gluons as well.
Nevertheless, to determine the behavior of $\bar\psi\psi$ in full 
QCD, the complete QCD path 
integral has to be solved, for which large-scale numerical lattice simulations are necessary. 
Another approach that has often been followed is to simplify the problem by working with 
effective degrees of freedom or by approximating the theory by a model that 
e.g.\ captures the correct symmetries. 

\subsection{Zero temperature}

The first results for the magnetic field-dependence of the quark condensate were obtained in 
such a model treatment, the Nambu-Jona-Lasinio (NJL) model~\cite{Schramm:1991ex,Gusynin:1995nb}. 
In particular, it was found that the magnetic field always enhances 
the condensate, a mechanism that was dubbed {\it magnetic catalysis}~\cite{Gusynin:1995nb}. 
This result was argued to 
be universal and model-independent. Indeed, $\bar\psi\psi$ was found to undergo magnetic catalysis 
in a host of different model and effective theory frameworks, see the recent review~\cite{Shovkovy:2012zn} 
and further references therein. 
The basic ingredient for magnetic catalysis was argued to lie in the dimensional reduction of the 
system for very strong magnetic fields~\cite{Gusynin:1995nb}. The energy of the zeroth Landau level is $B$-independent 
(cf.\ Eq.~(\ref{eq:contland})), whereas the next level lies much higher if $B$ is large, and its contribution 
to most physical observables is thus suppressed. 
In addition, the lowest Landau level effectively describes a one-dimensional system, since only the 
longitudinal momentum $p_z$ appears in it. This reduced dimensionality -- together 
with the fact that the degeneracy of the levels is proportional to $B$ -- leads to an increased phase space 
at low energies and thus supports the condensation of low eigenvalues. In turn, 
through the Banks-Casher relation~\cite{Banks:1979yr}, 
an increased density of the low eigenvalues translates to an 
enhancement of the quark condensate. 
Notice that this mechanism operates at high magnetic fields and is thus 
complementary to the arguments about the $\mathcal{O}(B^2)$ dependence of 
the condensate discussed in Eq.~(\ref{eq:correspondence}).

\begin{figure}[ht!]
\centering
\vspace*{-.3cm}
\mbox{
\includegraphics*[width=7.3cm]{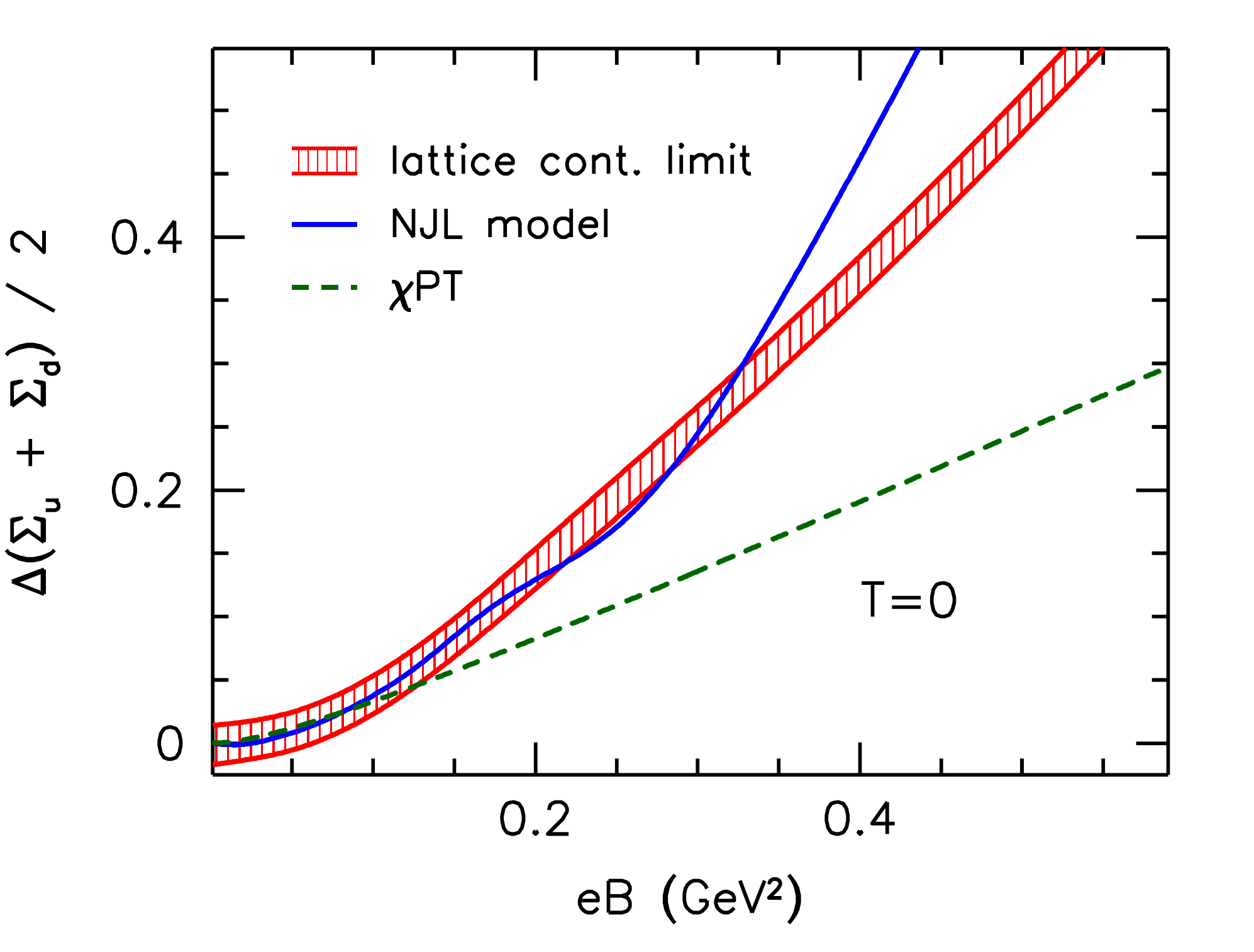} \quad
\includegraphics*[width=7.3cm]{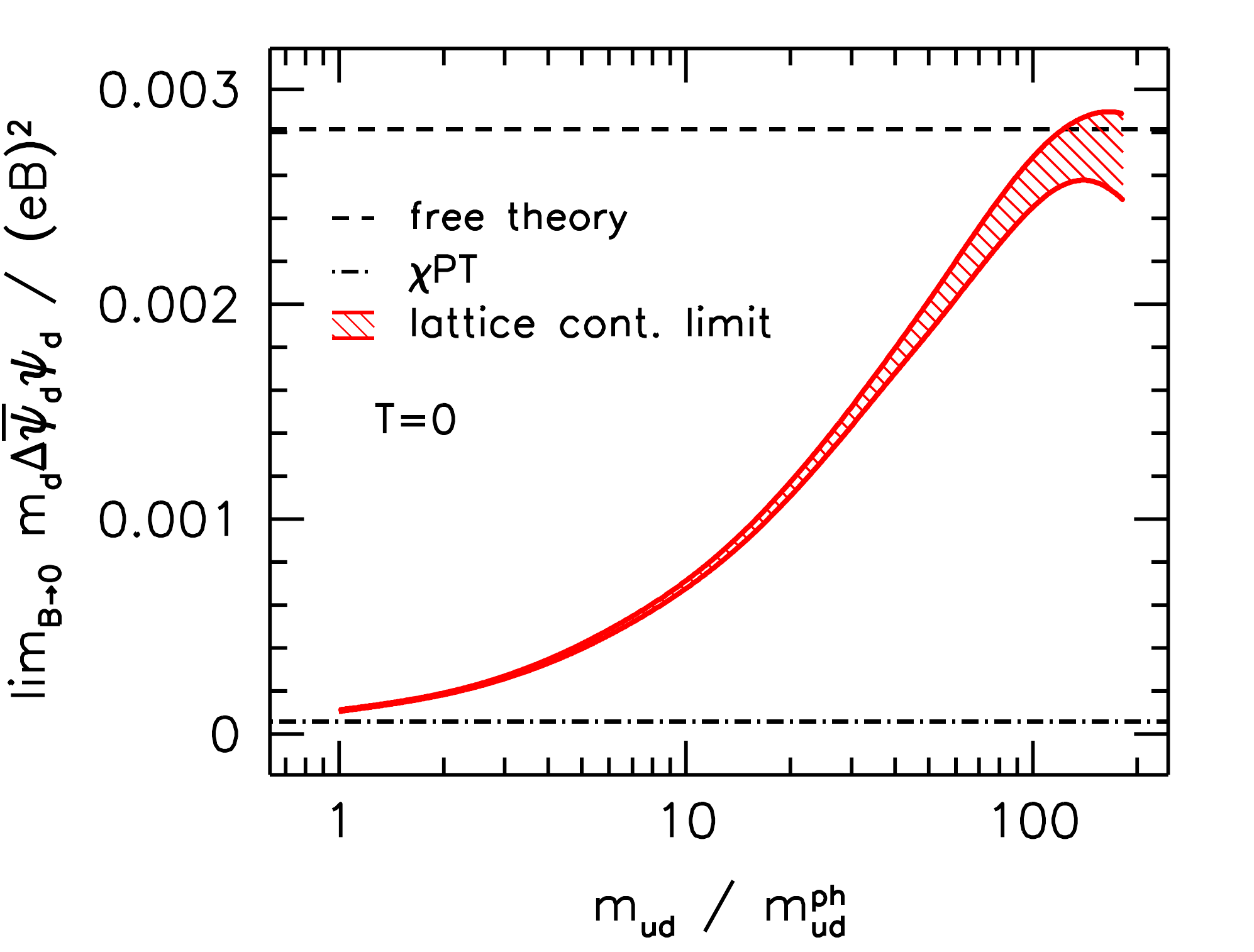} }
\vspace*{-.5cm}
\caption{\label{fig:pbpT0}Left panel: the continuum limit of the lattice results for change of the condensate, compared to 
$\chi$PT~\protect\cite{Cohen:2007bt} and to the NJL model prediction~\protect\cite{Gatto:2010pt}.
Right panel: the dependence of the coefficient of the quadratic contribution to the condensate, as a function of the quark mass. 
The continuum extrapolated lattice results interpolate between the limiting values, given by the free case 
($m\to\infty$) and by $\chi$PT ($m\to0$).}
\vspace*{-.2cm}
\end{figure}

After the first model studies, numerical lattice simulations have also been performed to investigate the effect 
of magnetic fields on the condensate, see the review~\cite{D'Elia:2012tr}. 
Magnetic catalysis was confirmed in the quenched 
theory~\cite{Buividovich:2008wf}, as well as in dynamical simulations with larger-than-physical pion 
masses with three~\cite{D'Elia:2010nq,D'Elia:2011zu} and with two colors~\cite{Ilgenfritz:2012fw}. 
In Refs.~\cite{Bali:2011qj,Bali:2012zg} we simulated full QCD with staggered quarks at physical masses and also 
performed the extrapolation of the results to the continuum limit. The average of the up and down quark condensates at $T=0$ is plotted in the left panel of Fig.~\ref{fig:pbpT0} 
and also compared to chiral perturbation theory ($\chi$PT) and to the NJL model, 
showing a qualitative -- and, for low $B$, even quantitative -- agreement. In summary, it is safe to say that at zero temperature all 
approaches agree and give a quark condensate that monotonously increases as $B$ grows. 

To quantify the strength of magnetic catalysis, in Ref.~\cite{Bali:2014kia} we 
also determined the coefficient of the quadratic enhancement of $\bar\psi\psi(B)$ for various values of the quark masses. 
On the one hand, for asymptotically high masses quarks and gluons decouple and the condensate can be calculated neglecting gluonic interactions 
(see App.~\ref{sec:app1}). On the other hand, if the mass approaches zero, $\chi$PT may be employed. Then, to leading order the condensate is 
calculated assuming free charged pions as effective degrees of freedom. The corresponding coefficient is again 
related to the $\beta$-function, but since pions are spinless, this time the {\it scalar} QED $\beta$-function coefficient 
$\beta_1^{\rm scalar}=1/(48\pi^2)$ appears. In the two limits we get~\cite{Endrodi:2013cs,Bali:2014kia}
\be
\lim_{B\to0} \;m\cdot \frac{\Delta \bar\psi\psi}{(eB)^2} = 
(q/e)^2\cdot
\begin{cases}
N_c\cdot \beta_1 , & m\to \infty \\
\beta_1^{\rm scalar} /4,  & m\to0,
\end{cases}
\label{eq:pbpcoeff}
\ee
where $N_c=3$ denotes the number of colors. 
As visible in the right panel of Fig.~\ref{fig:pbpT0}, the lattice results 
are completely consistent with this expectation and 
smoothly interpolate between the two extremes as the quark mass is varied. 

\subsection{Nonzero temperature}

Most of the above mentioned models and effective theories predicted 
magnetic catalysis to be dominant for all temperatures $T$, both in the confined and in the deconfined phase of QCD. 
This uniform enhancement of the condensate also implied that the magnetic field shifts the restoration of chiral symmetry 
to higher temperatures, i.e.\ that $T_c(B)$ {\it increases}. Results obtained within various different frameworks 
seemed to point towards this conclusion, see the reviews~\cite{Fraga:2012rr,*Gatto:2012sp}. In addition, lattice simulations 
with larger-than-physical quark masses also supported this picture~\cite{D'Elia:2010nq,Ilgenfritz:2012fw}. 

\begin{figure}[ht!]
\centering
\vspace*{-.1cm}
\mbox{
\includegraphics*[width=7.3cm]{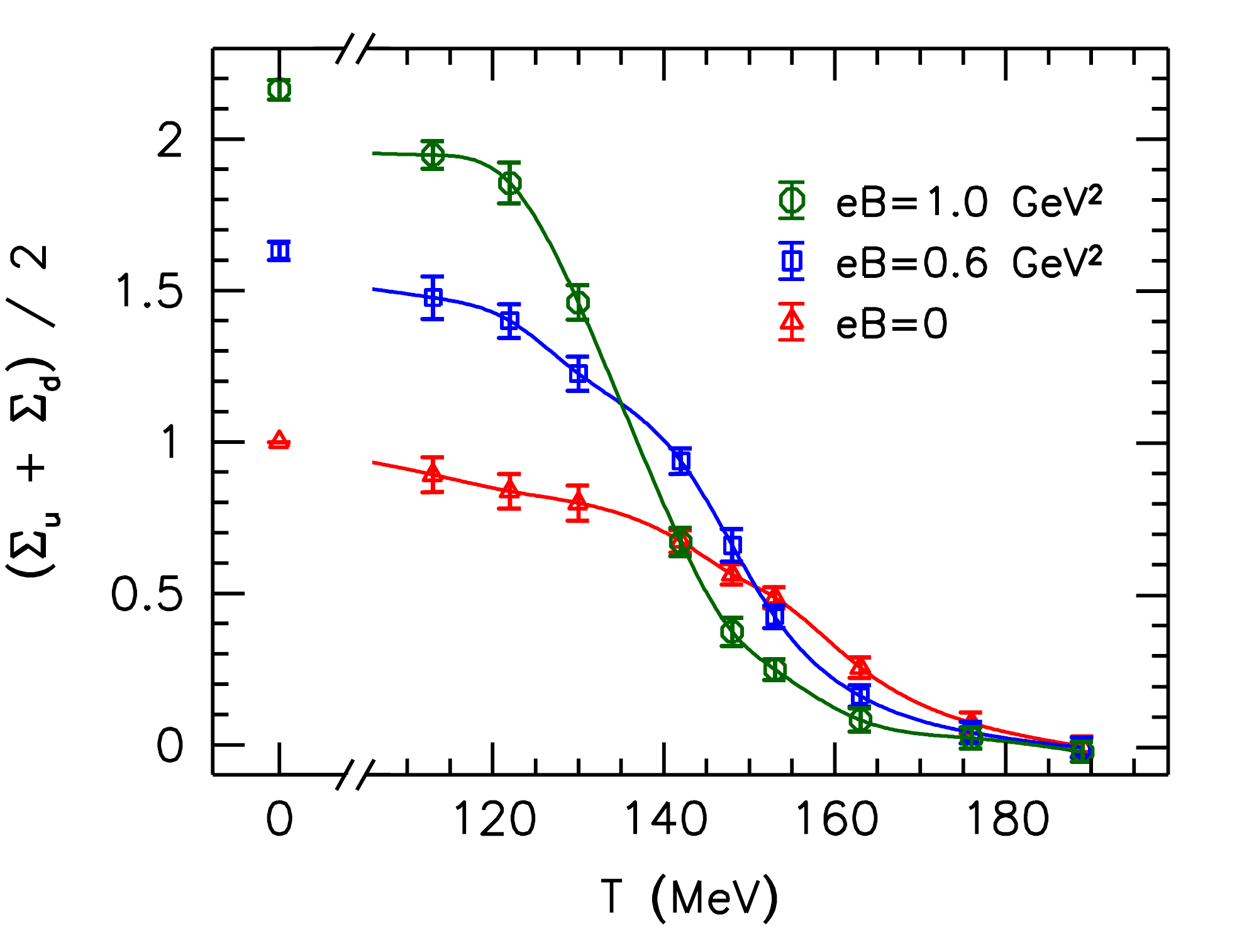} \quad
\includegraphics*[width=7.25cm]{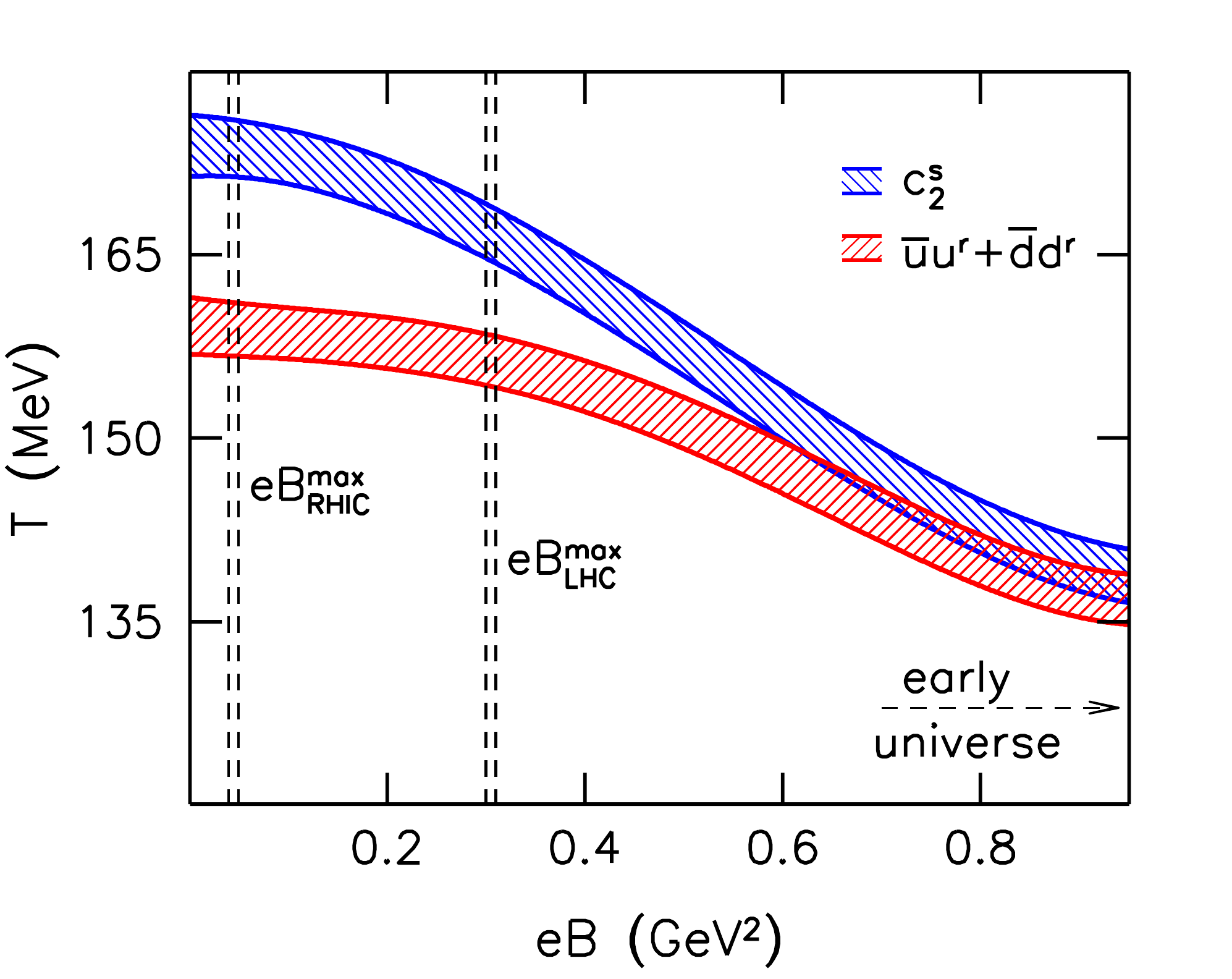} }
\vspace*{-.5cm}
\caption{\label{fig:pd}Left panel: magnetic catalysis at low temperatures and inverse magnetic catalysis in the transition region. As a result, 
the transition temperature -- identified by the inflection point of the condensate -- is decreased by $B$.
Right panel: the QCD phase diagram in the $B-T$ plane, with $T_c$ defined as the inflection point of the quark condensate 
(red band) and as that of the strange quark number susceptibility (blue band). }
\end{figure}

In contrast, in the large scale study in Refs.~\cite{Bali:2011qj,Bali:2012zg} employing staggered quarks with physical masses 
and a continuum 
extrapolation, we found that magnetic catalysis gets weaker as the temperature 
increases, and in the transition region the condensate is even reduced by the magnetic field, see 
the left panel of Fig.~\ref{fig:pd}. 
This behavior -- that we dubbed 
{\it inverse magnetic catalysis} -- also implied that chiral symmetry restoration occurs at lower temperatures as 
$B$ grows, i.e.\ that $T_c(B)$ {\it decreases}. The resulting phase diagram is plotted in the right panel of Fig.~\ref{fig:pd}. 

In Ref.~\cite{Bali:2011qj} we identified two reasons why inverse catalysis 
was not observed in earlier lattice simulations~\cite{D'Elia:2010nq}: large cutoff effects and larger-than-physical 
quark masses. In fact, we explicitly demonstrated that inverse catalysis is only active for the light flavors and disappears 
for the heavier strange quark. In a follow-up paper~\cite{Bruckmann:2013oba} we discussed the possible mechanisms behind inverse 
catalysis. It turns out that the magnetic field induces two different contributions to 
the condensate: one that originates 
from the direct interaction between $B$ and valence quarks, and
one that stems from the indirect coupling between $B$ and 
gluons~\cite{D'Elia:2011zu}. 
In the path integral language, the former corresponds to the operator insertion, whereas the latter 
to the fermionic determinant. 
Separating these contributions revealed that the direct term always enhances the condensate, while the indirect contribution 
reduces it in the transition region. If the quark mass is small, the indirect effect turns out to dominate and induces inverse catalysis around $T_c$~\cite{Bruckmann:2013oba}. From the perturbative point of view, the indirect effect may be described by the coupling of 
the background magnetic field (i.e.\ external photons) to virtual sea quark loops that interact with gluons. Due to the sea quark propagators, this diagram is proportional to $1/m^2$, revealing why it can only dominate for light quarks.

\begin{wrapfigure}{r}{7.cm}
 \centering
 \vspace*{-.3cm}
\includegraphics[width=7.cm]{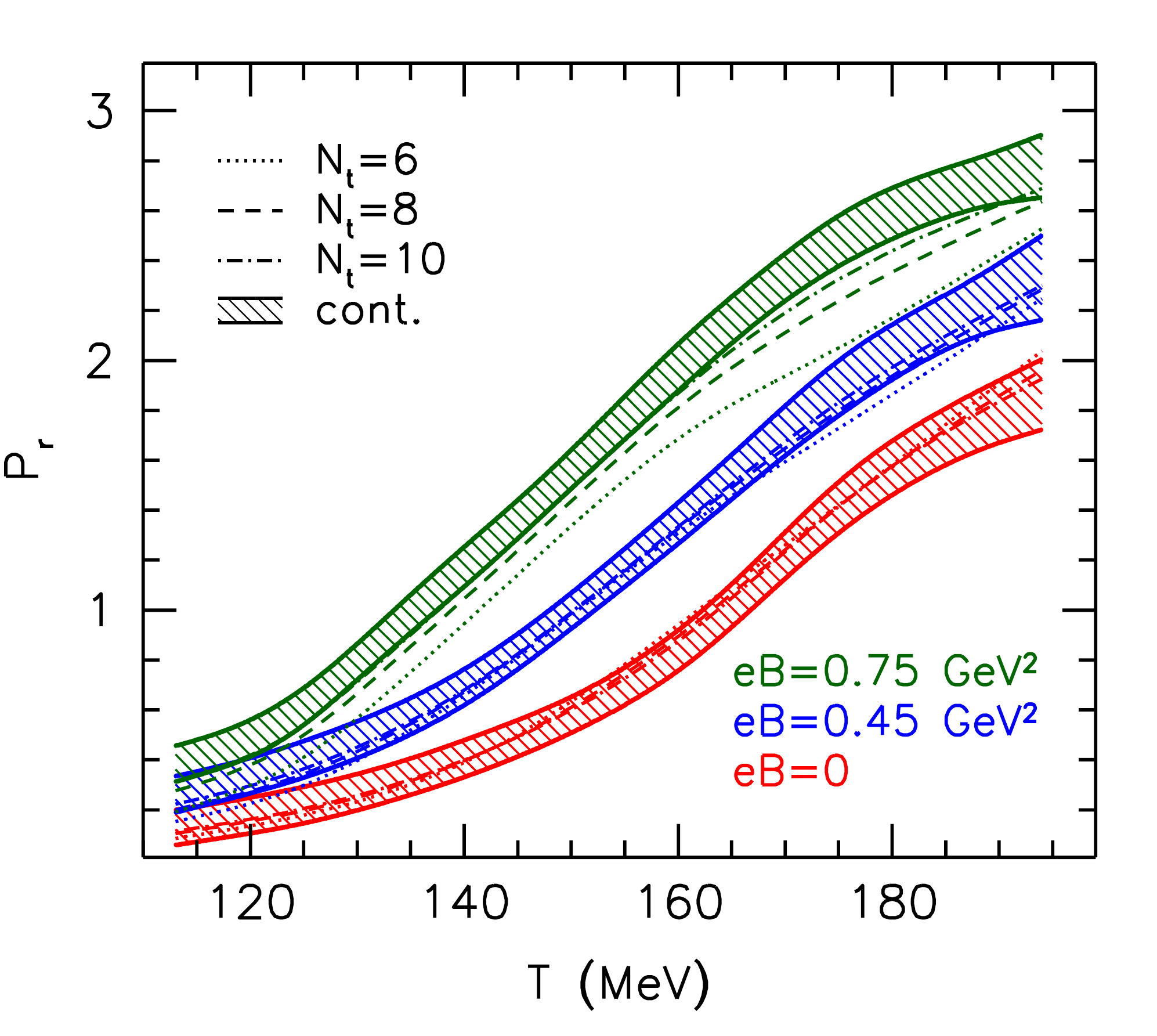}
\vspace*{-.65cm}
 \caption{\label{fig:ploop}The Polyakov loop as a function of the temperature, for various values of $B$.}
\vspace*{-.5cm}
\end{wrapfigure}

We then proceeded to identify the relevant gluonic degrees of freedom that this
indirect effect couples to. The most important gauge degree of freedom around $T_c$ is the Polyakov loop,
the path-ordered parallel transport 
$P = \Tr\, \mathcal{P}\exp (\int_0^{1/T} \!A_4 \,dt)$
winding around the temporal direction of the lattice.
Its dependence on the temperature and on the magnetic field is shown in Fig.~\ref{fig:ploop}.
The results show 
that $P$ is drastically enhanced by the magnetic field in the transition region. 
This implies that the inflection point of $P$ is
shifted to lower temperatures, revealing that around $T_c$ the magnetic 
field favors gauge configurations 
corresponding to the deconfined phase, i.e.\ with small condensate. 
This leads eventually to the inverse catalysis of $\bar\psi\psi$ in the transition region.

In addition, we have recently determined various thermodynamic observables related to the QCD 
equation of state~\cite{Bali:2014kia} and found that these also support the reduction of $T_c$ by the magnetic field.
Lattice results indicating this tendency have since been obtained using the overlap quark discretization 
in $N_f=2$ QCD~\cite{Bornyakov:2013eya} and in two-color QCD with four equally charged staggered 
quark flavors~\cite{Ilgenfritz:2013ara}. 
Besides the relevance of these findings for the physics of off-central heavy-ion 
collisions and early universe cosmology, the results have been highly useful 
for improving low-energy models of QCD. After the first attempts 
using simple models like the bag model~\cite{Fraga:2012fs} or large 
$N_c$ arguments~\cite{Fraga:2012ev}, 
there is extensive ongoing work to implement inverse catalysis in different 
frameworks, e.g.\ by tuning the free model parameters to fit the lattice 
results or by looking for different mechanisms
~\cite{Fukushima:2012kc,*Chao:2013qpa,
*Fraga:2013ova,*Ferreira:2013tba,*Kamikado:2013pya,*Ferrer:2014qka,*Fayazbakhsh:2014mca,*Farias:2014eca,*Ferreira:2014kpa,*Ayala:2014iba,*Ayala:2014gwa,*Andersen:2014oaa}.

\section{Summary}

In the first part of the talk, I discussed the lattice Dirac eigenvalues of 
free quarks exposed to a background magnetic field in two dimensions. 
The spectrum -- after a simple 
shift in the magnetic field -- coincides with Hofstadter's butterfly. 
Although the butterfly is a mere lattice artefact and is eliminated in the 
continuum limit, the low-$B$ behavior of the eigenvalues does represent 
continuum physics. In particular, I derived the correspondence~(\ref{eq:correspondence}) 
between three, seemingly independent phenomena:
a) the structure of the wings of Hofstadter's butterfly at small $B$,
b) the absence of asymptotic freedom in QED and 
c) the leading-order magnetic catalysis of the quark condensate.

In the second part I turned to full QCD in four dimensions. Towards the chiral limit, 
chiral perturbation theory can be used to relate the $\mathcal{O}(B^2)$ magnetic 
catalysis to the positivity of the QED $\beta$-function 
(in this case the scalar QED $\beta$-function coefficient appears). 
Continuum extrapolated lattice results also confirm this scenario 
at zero temperature. 
For temperatures around $T_c$, however, 
gluonic interactions tend to 
reduce the condensate and induce inverse magnetic catalysis. This mechanism operates 
via the indirect interaction between the magnetic field and the gauge degrees of freedom 
(most importantly, the Polyakov loop) via charged sea quark loops.
As a result, the transition temperature decreases as $B$ grows and the phase diagram 
looks as depicted on the right panel of Fig.~\ref{fig:pd}. 
Progress has been made recently to implement such an indirect coupling in 
low-energy models of QCD.

\appendix
\section{Condensate in magnetic fields}
\label{sec:app1}

To reveal the role played by the QED $\beta$-function in the leading 
magnetic field-dependence of the condensate, it is instructive to first consider 
the free energy density $f$ of the system. 
To enable a direct comparison between the four- and the two-dimensional cases, 
the system dimension $2\le D\le4$ is not yet specified. 
At zero temperature, the logarithm of the partition function,
\be
\log\Z = \frac{1}{2} \sumint_{\;D} \log  \lambda_D^2,
\ee
is written in terms of the $D$-dimensional energy eigenvalues 
and degeneracies
\be
\lambda_D^2 = m^2 + 2qB(n+1/2-s) + \sum_{i=3}^D p_i^2, \quad\quad\quad
\sumint_{\;D} = 2\cdot\frac{\Phi}{2\pi} \,\sum_{n=0}^\infty \;\sum_{s=\pm1/2} \;\,
\prod_{i=3}^D L_i \int \frac{\d p_i}{2\pi},
\ee
where $\Phi=qB L_xL_y$ is the flux of the magnetic field. 
To regularize the sums and integrals the $\zeta$-function regularization 
(Mellin transform) is 
used\footnote{I would like to thank Falk Bruckmann for the countless discussions that we had 
about this and similar calculations.},
\be
\log \lambda_D^2 = -\left.\frac{\partial}{\partial \alpha}\right|_{\alpha=0^+}
(\lambda_D^2)^{-\alpha}
, \quad\quad\quad
(\lambda_D^2)^{-\alpha}   = 
\frac{1}{\Gamma(\alpha)} 
\int_0^{\infty}\d z \,z^{\alpha-1} \,e^{-\lambda_D^2\,z}, \quad\quad \textmd{Re } \lambda_D^2 > 0.
\ee
Inserting this regularization, all sums and integrals can be
performed. Differentiating with respect to $\alpha$,
\be
\left.\frac{\partial}{\partial \alpha}\right|_{\alpha=0^+} \frac{z^\alpha}{\Gamma(\alpha)} = 1,
\ee
results in Schwinger's proper time 
representation~\cite{Schwinger:1951nm} for the free energy density,
\be
f \equiv -\frac{1}{V_D}\log\Z = 
\frac{2qB}{(4\pi)^{D/2}} \int_0^\infty \!\frac{\d z}{z^{D/2}} \,e^{-m^2 z} \,\coth(qBz),
\label{eq:fD2}
\ee
where $V_D=\prod_i L_i$ is the $D$-dimensional volume. 

To obtain the leading dependence on the magnetic field, $\Delta f\equiv f(B)-f(B=0)$ is expanded in $qB$,
\be
\Delta f = \frac{(qB)^2}{m^{4-D}} \, \frac{2}{3(4\pi)^{D/2}} \,\Gamma(2-D/2) + \mathcal{O}(B^4).
\ee
The $\mathcal{O}(B^2)$ term in the free energy density corresponds to the 
coupling of a fermion loop to two external photon legs (the external photons 
represent the background magnetic field), i.e.\ the photon vacuum polarization 
diagram~\cite{Schwinger:1951nm,Abbott:1981ke,Dunne:2004nc}. 
In $D=4$ dimensions, this diagram is logarithmically divergent and its coefficient 
equals the lowest-order QED $\beta$-function coefficient $\beta_1=1/(12\pi^2)$. 
In $D=2$ dimensions, the vacuum polarization is ultraviolet finite, but 
the coefficient is still proportional to $\beta_1$,
\be
\Delta f = \frac{(qB)^2}{2} \cdot \beta_1 \cdot
\begin{cases}
 \frac{1}{\epsilon} -\gamma -\log m^2 +\mathcal{O}(\epsilon), & D=4-2\epsilon \\
 4\pi / m^2, & D=2
\end{cases}
\quad + \, \mathcal{O}(B^4),
\label{eq:fD}
\ee
where $\gamma$ is Euler's constant.

Differentiating Eq.~(\ref{eq:fD2}) with respect to the mass gives the 
condensate,
\be
\bar\psi\psi \equiv -\frac{\partial f}{\partial m}  
= \frac{4mqB}{(4\pi)^{D/2}} \int_0^\infty \!\frac{\d z}{z^{D/2-1}} \,e^{-m^2 z} \,\coth(qBz),
\label{eq:pbpproper}
\ee
which, after a similar expansion in the magnetic field as above, gives
\be
\Delta \bar\psi\psi = -\frac{\partial \Delta f}{\partial m} = 
(qB)^2\cdot \beta_1 \cdot
\begin{cases}
 1/m, & D=4 \\
 4\pi/m^3, & D=2
\end{cases}
\quad +\, \mathcal{O}(B^4).
\label{eq:pbpb1}
\ee
Notice that the difference between $D=4$ and $D=2$ is the different power of 
the mass to account for the dimensionality, and the factor $4\pi$ that 
stems from the additional phase space in the third and fourth dimensions. 
Altogether, Eq.~(\ref{eq:pbpb1}) shows that the positivity of $\beta_1$ -- i.e., 
the absence of asymptotic freedom in QED -- results in an $\mathcal{O}(B^2)$
increase of the condensate of free quarks, both in two and in 
four dimensions. This argument has been used in Refs.~\cite{Endrodi:2013cs,Bali:2013txa,Bali:2014kia} 
to relate magnetic catalysis to the positivity of the QED 
$\beta$-function for $D=4$. 
Note that for quarks the number $N_c$ of colors also appears in Eq.~(\ref{eq:pbpb1}) 
as a multiplicative factor.

\bibliographystyle{JHEP_notitle_mcite}
\bibliography{butterfly}

\end{document}